\documentclass[iop]{emulateapj}

\newcommand{\eqref}[1]{(\ref{#1})}

\def\lesssim{\mathrel{\hbox{\rlap{\hbox{\lower4pt\hbox{$\sim$}}}\hbox{$<$}}}}
\def\gtrsim{\mathrel{\hbox{\rlap{\hbox{\lower4pt\hbox{$\sim$}}}\hbox{$>$}}}}

\shorttitle{HL Tau}
\shortauthors{Akiyama et al.}

\usepackage{color}
\usepackage{ulem}
\begin{document}


\title{PLANETARY SYSTEM FORMATION IN PROTOPLANETARY DISK AROUND HL TAURI}


\author{EIJI AKIYAMA\altaffilmark{1}, YASUHIRO HASEGAWA\altaffilmark{1,2}, MASAHIKO HAYASHI\altaffilmark{1}, and SATORU IGUCHI\altaffilmark{1,3}}

\affil{\altaffilmark{1}National Astronomical Observatory of Japan, 2-21-1, Osawa, Mitaka, Tokyo, 181-8588, Japan; eiji.akiyama@nao.ac.jp, yasuhiro.hasegawa@nao.ac.jp \\
\altaffilmark{3}School of Mathematical and Physical Science, SOKENDAI (The Graduate University for Advanced Studies), Hayama, Kanagawa 240-0193, Japan
\\
}
\altaffiltext{2}{EACOA fellow}




\begin{abstract}
We re-process the Atacama Large Millimeter/Submillimeter Array (ALMA) long-baseline science verification data taken toward HL Tauri. As shown by the previous work, we confirm that the high spatial resolution ($\sim$ 0\farcs019, corresponding to $\sim$ 2.7 AU) dust continuum images at $\lambda = 0.87$, $1.3$, and $2.9$ mm exhibit a multiple ring-like gap structure in the circumstellar disk. Assuming that the observed gaps are opened up by currently forming, unseen bodies, we estimate the mass of such hypothetical bodies based on following two approaches; the Hill radius analysis and a more elaborated approach developed from the angular momentum transfer analysis in gas disks. For the former, the measured gap widths are used for calibrating the mass of the bodies, while for the latter, the measured gap depths are utilized. We show that their masses are likely comparable to or less than the mass of Jovian planets, and then discuss an origin of the observed gap structure. By evaluating Toomre's gravitational instability (GI) condition and cooling effect, we find that the GI might be a possible mechanism to form the bodies in the outer region of the disk. As the disk might be gravitationally unstable only in the outer region of the disk, inward planetary migration would be needed to construct the current architecture of the hypothetical bodies. We estimate the gap-opening mass and show that type II migration might be able to play such a role. Combining GIs with inward migration, we conjecture that all of the observed gaps may be a consequence of bodies that might have originally formed at the outer part of the disk, and have subsequently migrated to the current locations. While ALMA's unprecedented high spatial resolution observations can revolutionize our picture of planet formation, more dedicated observational and theoretical studies are needed in order to fully understand the HL Tau images.
\end{abstract}






\section{Introduction}
Explorations and studies of young circumstellar disks reveal important physical processes that are fundamental to the formation and evolution of planetary systems. The spatial distribution of disk materials around young stellar objects (YSOs) is a direct source of information from observations. By observing the distribution at different evolutional stages, one can investigate where and how the disk materials will be used for the formation of planetary systems. A number of interesting structures have been observed in circumstellar disks around low and intermediate mass YSOs at optical, infrared, and radio wavelengths \citep{thalmann10,thalmann14,andrews11,hashimoto12,debes13,fukagawa13,grady13,isella13,akiyama15}. These structures are gaps, spiral arms, cavities, and asymmetry, and imply the presence of planets that may be fully formed already or may currently be forming. Since the recent radio observations have provided only large-scale structures of disks due to low spatial resolutions, the advent of Atacama Large Millimeter/Submillimeter Array (ALMA) with significantly upgraded capabilities is indeed demanded that are currently providing an invaluable opportunity to probe the disk interior with high spatial resolutions and sensitivities.

HL Tauri (hereafter HL Tau) has been intensively investigated observationally over the last decades, because of its nearness to the Sun with a distance of approximately 140 pc \citep{elias78,kenyon94}. It has been classified as an in-between Class I and II T Tauri star based on its flat spectrum for a range of $2 \mu \mbox{m} \la \lambda \la 60 \mu \mbox{m}$ in the wavelengths in the spectral energy distribution (SED) \citep{menshchikov99}. The classification is supported by a number of additional observational features, such as molecular outflow \citep{monin96}, large mass accretion rates of $0.9 \times 10^{-5} \ M_\sun \ {\rm yr}^{-1}$ onto the central star \citep{hayashi93}, and a circumstellar envelope with infall or rotational motion which is traced by the CO emission \citep{sargent91,hayashi93}. More recently, \citet{partnership} have summarized the results of the ALMA long-baseline science verification campaign, and have shown that the dust continuum emissions at 0.87, 1.3, and 2.9  mm surprisingly exhibit a multiple gap structure in the circumstellar disk around HL Tau. The astonishing ALMA images have immediately stimulated a number of follow-up studies, wherein origins of observed gaps are discussed. While condensation fronts (known as snow lines) of certain molecule \citep[e.g.,][]{zhang15}, and/or secular gravitational instability \citep[e.g.,][]{ward00,youdin11,takahashi14} may be one possibility, the most intriguing hypothesis would be a planet origin \citep[e.g.,][]{tamayo15,partnership,dipierro15}. In fact, the ALMA images provide some signatures that potentially imply ongoing planet formation in the disk, such that offset of the central position between continuum peaks and the center of the rings increases with the distance \citep{partnership}, and that a gap at 25 AU from the central star can be curved by a planet of approximately 0.5 $M_J$ \citep{kanagawa15}. 

In this paper, we explore the planet hypothesis further, and discuss how plausible the current theory of planet formation is in understanding the observed gap structure. To proceed, we re-process the ALMA archive data because a better quality of the dust continuum maps is desired to estimate the mass of hypothetical unseen planets more accurately. We find that there is a potential to improve the quality of the images that were previously released for the public. This will become possible, because only a small number of long baseline components have been included in the ALMA observations toward HL Tau. As shown by \citet{partnership}, we confirm that re-processed surface brightness presents multiple dips at 13.5, 32.4, 65.2, and 77.2 AU. Assuming that these gaps are formed by unseen, massive objects, we find that at least two planets in the outer region of the disk are very likely to be formed by gravitational instability (GI), while two planets currently located in the inner region of the disk may be transported from the outer part of the disk by type II migration. Thus, the ALMA images of HL Tau can provide profound insights into the current theory of planet formation.

\section{Observation and Data Reduction}
\label{obs}
As a part of the commissioning for the extended configuration including 10 km-long baseline, HL Tau was selected as a science verification target and observed on October 24th 2014 at band 3, 6, 7, corresponding to 2.9, 1.3, and 0.87 mm in wavelength, respectively, in the program ADS/JAO.ALMA\#2011.0.00015.SV \citep[see also][]{partnership}. All of the data including observed raw visibility data, calibrated data set, and images are available in the ALMA science verification data archive. The baseline lengths were between $\sim$ 15 and $\sim$ 15238 m, providing a synthesized beam of 0\farcs0822 $\times$ 0\farcs0544 at band 3, 0\farcs0346 $\times$ 0\farcs0225 at band 6, and 0\farcs0189 $\times$ 0\farcs0285 at band 7. The fields of view for bands 3, 6, and 7 were $\sim$ 60\farcs7, $\sim$ 26\farcs5, and $\sim$ 18\farcs0, respectively, in FWHM for the 12 m array. The number of antennas used in the 12 m array was between 28 and 37. The total integration time including overheads and calibration is 14.5 h in band 3, 9.5 h in band 6, and 11.9 h in band 7 and the corresponding integration time on source for each band after data reduction is 7.2 h, 4.9 h, and 5.4 h, respectively.  

The ALMA calibration includes simultaneous observations of the 183 GHz water line with water vapor radiometers that measure the water column in the antenna beam, later used to reduce the atmospheric phase noise. J0510+1800, J0423-0120, J0238+1636, and J0237+2848 were used to calibrate the bandpass and amplitude, and the quasars J0431+2037 and J0431+1731 were selected as phase calibrators in the observations. 

Data reduction was performed using version 4.2.2 of the Common Astronomy Software Applications package (CASA) and the deconvolution was done using task CLEAN to make images from the final calibrated visibilities. The continuum data itself as a model was used in self-calibration and only phase correction was applied. The self-calibration was conducted in four steps by decreasing the solution interval from 600 sec, 300 sec, 150 sec, and finally to scan time. To more clearly and accurately recover the extended component, we increase the quantity of contributing data by modifying, for example, the minimum number of baselines per antenna (minblperant) and the minimum signal to noise ratio (minsnr) required for gain solutions during the self-calibration. We applied minblperant = 4 and minsnr = 2 for the self calibration while minblperant = 6 and minsnr = 3 are recommended in the ALMA CASA guide. Since some structures are already visible even in the dirty image, we at first selected the only central emission component that can be assumed as a Gaussian  distribution for CLEAN process. Then, we extended the CLEAN box area gradually, and finally we CLEANed all of the components.  

\section{Results}
\begin{figure}[h]
\begin{center}
\includegraphics[scale=0.6]{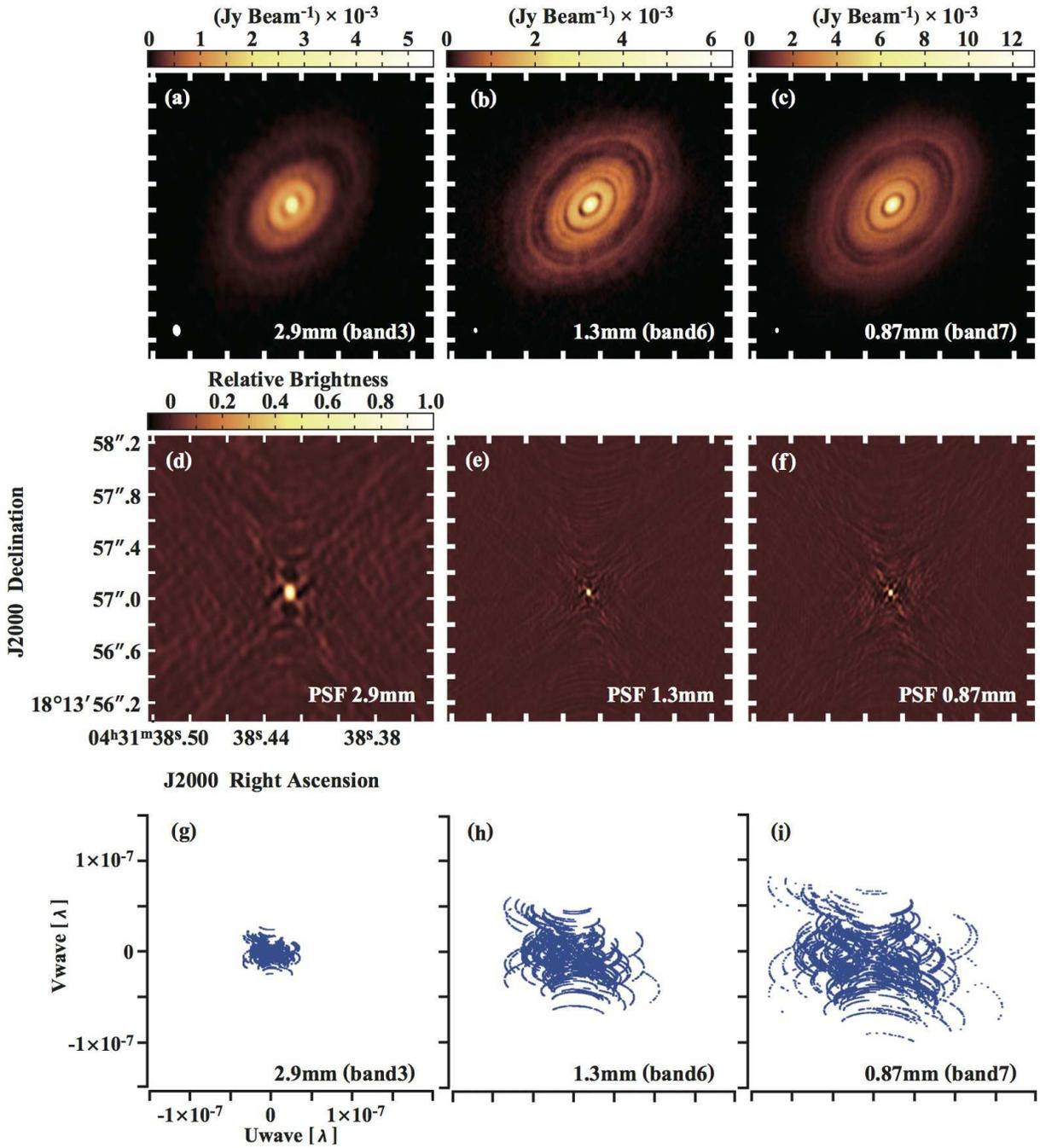}
\caption{Dust continuum emission maps of HL Tau at 2.9 mm (a), 1.3 mm (b), and 0.87 mm (c) are provided with corresponding images of their synthesized beam maps (psf) and uv-coverages. The beam size at each wavelength is provided as a white filled ellipse in the lower left corner in the top panels (A color version of this figure is available in the online journal).}
\label{fig:fig1}
\end{center}
\end{figure}

Figure \ref{fig:fig1} shows re-processed images of the dust continuum emission from the disk at each band (top panels), their corresponding synthesized beam maps (middle panels), and their uv-coverages (bottom panels). We have examined the surface brightness map by pixel by pixel along the direction of position angle of 138$\arcdeg$ \citep{partnership}, and identified the position of gaps where the surface brightness takes a minimum value. We have made use of the position to define the distance of the gaps  from the central star ($r_p$). As shown by \citet{partnership}, clear dips are seen at $\sim$ 13.5, $\sim$ 32.4, $\sim$ 65.2, and $\sim$ 77.2 AU from the continuum peak.

\begin{figure}[h]
\begin{center}
\includegraphics[scale=0.45]{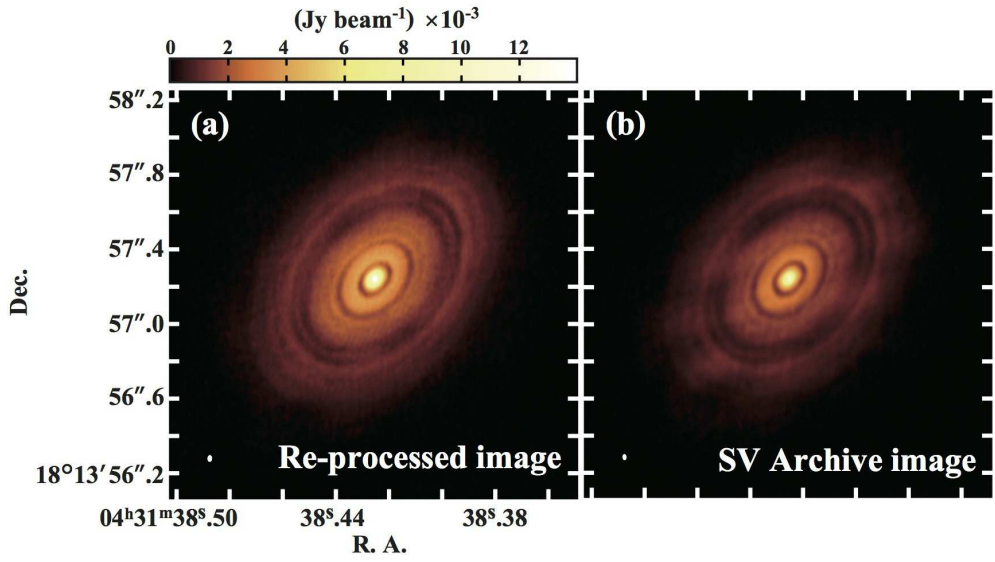}
\caption{Comparison of the 0.87 mm continuum dust emission of HL Tau between our re-processed image (a) and the archive data (b).   
(A color version of this figure is available in the online journal).}
\label{fig:fig2}
\end{center}
\end{figure}

Figure \ref{fig:fig2} shows the comparison between our re-processed 0.87 mm continuum image (a) and the original ALMA archive image (b). In the archive image in Figure \ref{fig:fig2} (b), radially-oriented fuzzy structures are seen especially from north-west to south-east direction, which are probably caused by missing short baseline components \citep[][private communication]{partnership}. In our re-processed images in Figure \ref{fig:fig2} (a), however, the structures are eliminated and the much clearer ring-like emission is retrieved in the outer region of the disk. One possible explanation for eliminating the structures can be the data flag condition we applied for the gain calibration. Since we increase the acceptable data during the self-calibration by reducing the minimum number of baselines per antenna and the minimum signal to noise ratio when determining the gain solution, the short spacing components were not flagged in the self-calibration. This also implies that the observation was conducted very accurately in the ALMA long baseline commissioning campaign since we accepted more data with low signal to noise ratio and as a result obtain the better image.

\begin{figure}[h]
\begin{center}
\includegraphics[scale=0.45]{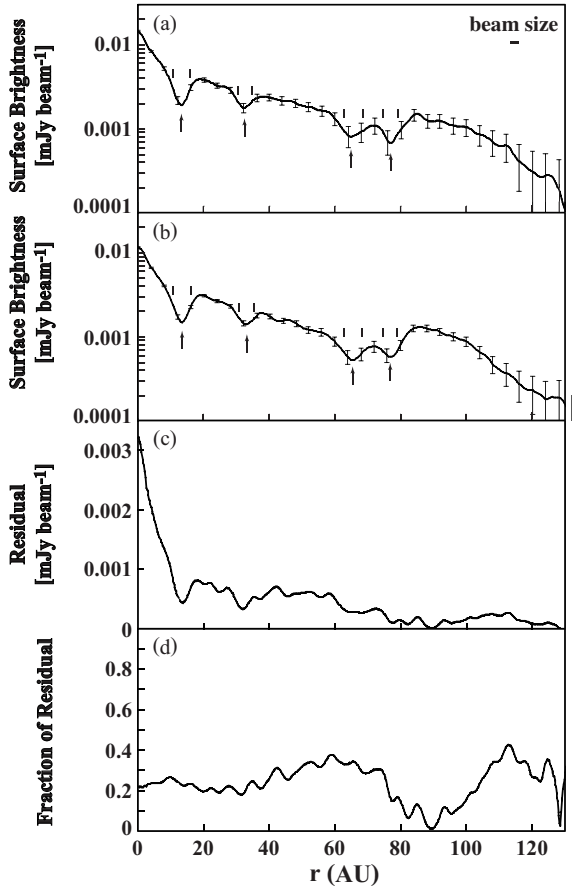}
\caption{
Panels (a) and (b) are the surface brightness profiles generated from our re-processed image (Figure \ref{fig:fig2} a) and the ALMA science verification archive data (Figure \ref{fig:fig2} b) along the semi-major axis (PA = 138$\arcdeg$), respectively. The error bars show 5 $\sigma$. Note that the systematic error in the absolute flux determination is expected to be approximately 10 \% at most (see ALMA Proposer’s Guide and Capabilities). The beam size is shown on the upper right corner of the panel (a) as a reference. The arrows indicate the location of each gap and pairs of the vertical solid lines describe the width at each gap. Panel (c) indicates the residual that is computed by subtracting (b) from (a), and the panel (d) represents the fraction of residuals with respect to (a).} 
\label{fig:fig3}
\end{center}
\end{figure}

Figure \ref{fig:fig3} a and b shows the radial profile of  the surface brightness along the semi-major axis \citep[P.A. = 138$\arcdeg$;][]{partnership}. We define the gap width ($\Delta r_p$) designated by sets of two vertical lines as follows; first, we measure the value of the surface brightness ($L_{gap}$) at the position of gaps ($r=r_p$) and the corresponding gap depth ($\Delta L_{gap}$) there. Second, we find out the positions at which the surface brightness takes the value of $L_{gap} + 0.5 \Delta L_{gap}$. As a result, we obtain the following gap widths; 5.0, 4.1, 6.2, and 4.5 AU, at $r_p$ = $\sim$ 13.5, $\sim$ 32.4, $\sim$ 65.2, and $\sim$ 77.2 AU in radius, respectively. Note that, while one may notice that there may be more ring-like structures in the continuum image of HL Tau, we focus on only four gaps whose widths are clearly specified. In other words, the resultant properties of gaps are measured accurately, so that our analysis can safely be applied (see below).

We also show the difference between our re-processed image and the ALMA archive one (see Figure \ref{fig:fig3}c). In the plot, the residual surface brightness is presented, by subtracting the profile of the ALMA archive image from the one generated from our reprocessed image. Figure \ref{fig:fig3}d finally represents the fraction of residuals with respect to the profile of our re-processed image. The results indicate that the difference becomes up to 20 to 40 \% in the gap region. The total flux densities at each band are measured to be 73 mJy in band 3, 800 mJy in band 6, and 2.3 Jy in band 7, and the corresponding achieved rms are down to 10, 18, and 24 $\mu$Jy, respectively. We thus confirm that our measurement is consistent with the flux density of 700 $\pm$ 10.3 mJy that is derived from the previous observations at $\lambda = 1.3$ mm taken by the Combined Array for Research in Millimeterwave Astronomy \citep[CARMA,][]{kwon11}. 

We have noticed through the re-processing of images that only 17 out of 780 antenna combinations had a baseline of more than 10 km. This corresponds to only two percent of the total number of baselines, and thus indicates that the images still have some potential to be improved. We quantify how much the long baseline components contribute to the images. This can be done by computing the maximum angular resolution ($\theta_{\rm max}$), which can be written as
\begin{equation}
\label{theta_max}
\theta_{\rm max} = a \times \frac{\lambda}{B_{\rm max}}, 
\end{equation}
where $\lambda$ is the wavelength, $B_{\rm max}$ is the maximum baseline length, and $a$ is a coefficient to calibrate how effective the longest baseline components are to create images. Note that $a$ should be between 0.5 and 1.0 if visibilities obtained by the longest baseline components are sufficiently contained in the data. For our re-processed images, $a$ becomes 1.6, and thus it can be concluded that images of, in particular, small structure still have some potential for further improvement in their angular resolution. While it is important to seek better images, the image quality achieved by our re-processing is sufficiently high for the following discussion, and we leave a more complete discussion of improving images for future work. 

\section{Implications for planet formation} 
\label{discu}

The dust continuum images with high spatial resolutions and sensitivities clearly show the gap morphology in the disk around HL Tau. Assuming that the gaps are opened up by unseen bodies, the precise measurements of gap widths make it possible to estimate the mass of the bodies more accurately than ever before. To this end, we utilize the observed radial surface brightness profile. Based on the mass estimate, we explore possible formation mechanisms of the bodies and discuss how plausible the current theories of planet formation are to account for the observed gap structure. We here focus on the four clear gaps as discussed above (see Figure \ref{fig:fig1}).

\subsection{Disk model}
\label{sec:model}

In this section, we describe a disk model that is utilized for estimating planet masses.

We adopt a simplified version of a disk model used in \citet{kwon11}; \citet{kwon11} consider the volume density of gas ($\rho_g(r, z)$) and the corresponding disk temperature ($T(r,z)$), both of which are functions of $r$ and $z$. On the other hand, we integrate $\rho_g((r,z))$ for the vertical direction and assume that the disk temperature can be modeled by the two characteristic temperatures ($T_s(r)$ and $T_i(r)$), where $T_s(r)$ represents the temperature of disk surfaces, and $T_i(r)$ denotes the temperature at the midplane. Thus, the surface density distribution of gas ($\Sigma_g(r)$) becomes essentially comparable to the famous similarity solution \citep{pringle81}, and can be written as
\begin{eqnarray}
 \label{eq:sigma}
 \Sigma_g(r) & = & \rho_0  \sqrt{\pi} H(r) \left( \frac{r}{r_c} \right)^{-p} \\ 
  & \times & \exp \left[ - \left( \frac{r}{r_c} \right)^{7/2-p-q/2} \right], \nonumber
\end{eqnarray}
where $\rho_0$ is a constant, $H(r)$ is the pressure scale height, $p$ is a fitting parameter, and $q=0.43$ is the exponent that involves the radial temperature profile (see Table \ref{table1}). Note that $\rho_0$ is computed for a given value of the disk mass ($M_{d}$). For both $T_s$ and $T_i$, a simple power-law distribution is assumed, that is, $T_s = T_{s0}(r/r_c)^{q}$ and $T_i = T_{i0}(r/r_c)^{q}$ following \citet{kwon11}. Since we currently examine sub-mm observations, $T_i$ plays an important role in reproducing the observations.

\begin{figure}[h]
\begin{center}
\includegraphics[scale=0.4]{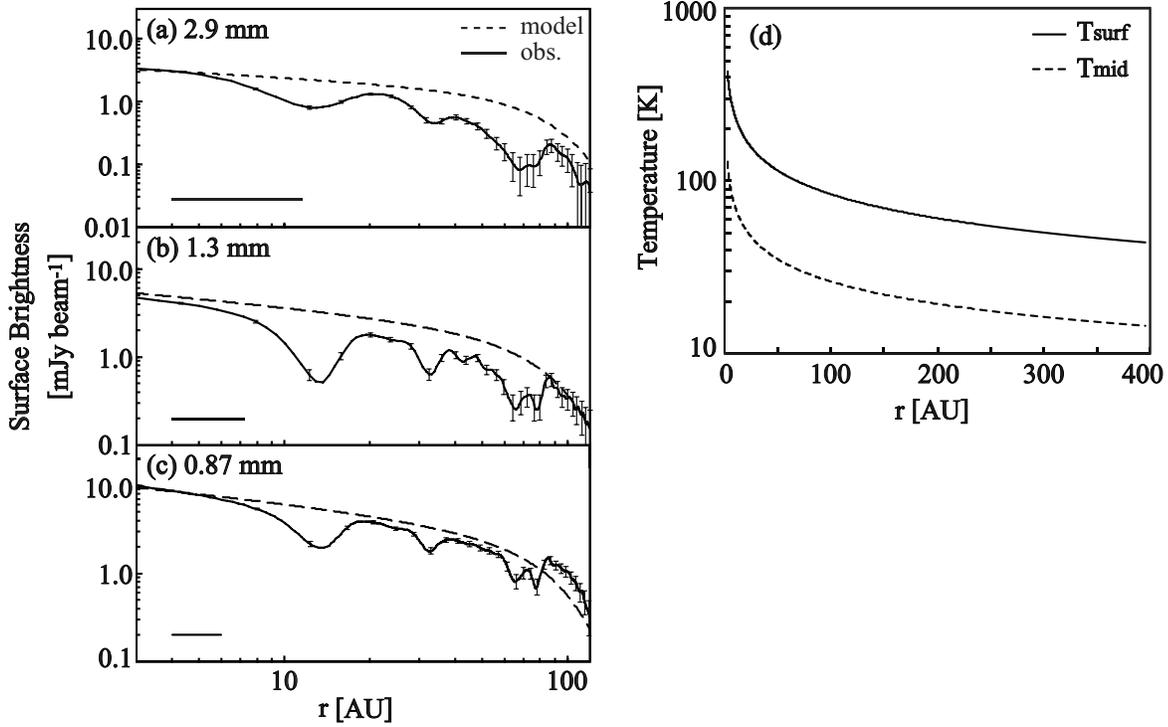}
\caption{
Panels (a), (b), and (c) show surface brightness profiles (see the solid curve) at 2.9, 1.3, and 0.87 mm in wavelengths along the semi-major axis (PA = 138$\arcdeg$), respectively. The dashed curve represents the profile generated by \citet{kwon11} model. The error bar represents 5 $\sigma$ and the horizontal line at the lower left corner indicates the beam size at each wavelength. Panel (d) represents the radial temperature distributions at the disk surface (solid curve) and the mid-plane (dashed curve) of the disk computed from \citet{kwon11} model.}
\label{fig:fig4}
\end{center}
\end{figure}

Based on the above disk model, \citet{kwon11} perform the Markov Chain Monte Carlo analysis, and derive a best set of model parameters with which the observed quantities such as images are well reproduced. We simply use the results of their best fit case (see Table \ref{table1}). One may then wonder how reasonable it is to interpret new ALMA observations using previous results derived from observations with moderate spatial resolutions. We examine this issue by calculating the surface brightness at each band to identify how well they match with the ALMA observations. Figure \ref{fig:fig4} shows the resultant surface brightness derived from Kwon's model. Note that for a given value of $\Sigma_g(r)$ and $T_i(r)$, the surface brightness was computed, following \cite{dullemond01} (see their equation (21)). The results indicate that the Kwon's model can reproduce the ALMA observations in all bands very well except for the gap regions. Thus, it can be concluded that disk models that can well reproduce the previous observations with moderate spatial resolutions are still valuable to characterize the global structure of disks.

\begin{table*}
\begin{minipage}{17cm}
\begin{center}
\caption{Summary of model parameters$^\dagger$}
\label{table1}
\begin{tabular}{cccccccccc}
\hline
$T_0^\ddagger$ & $p$ & $q$ & $\beta$ & $\kappa_{0, @230GHz}$ & $r_{in}$ & $r_c$ & gtd & $ M_{disk}$ & $i$ \\  
free param.& 1.064 & 0.43 & 0.729 & 0.01 &  2.4 AU & 78.9 AU & 100 & 0.1349 $M_\sun$ & 40.3 \\
\hline
\end{tabular}
\end{center}
$^\dagger$ see \citet{kwon11} and references therein.  \\
$^\ddagger$ We treat $T_0$  as a free parameter because the temperature can not be well constrained at mm/sub-mm wavelengths 
\citep[see Figure 3 in][]{kwon11}. 
\end{minipage}
\end{table*}

\subsection{The mass of hypothetical bodies}
\label{sec:hill}

We now estimate the mass of hypothetical bodies by which the observed gaps can be opened. To proceed, we make use of the surface brightness profile obtained in the above section. Note that it is assumed that the central star is located at the observed continuum peak. We estimate their masses by following two approaches; the Hill radius analysis and a more detailed method in which the balance of angular momentum transfer is considered. The advantage of the first method is simple and independent of disk quantities such as surface density, temperature, or turbulent viscosity, which are sometimes poorly constrained. The advantage of the second method is a good capturing of disk physics, wherein angular momentum transfer is triggered both by a planet and by gas disks due to viscosity and pressure.

First, we adopt the Hill radius analysis and examine the resultant mass of the bodies. Assuming that the gap width ($\Delta r_p$) is comparable to the Hill radius of the bodies, 
$\Delta r_p$ is related to the mass of the bodies ($M_{p,hill}$) as 
\begin{equation}
 \Delta r_p \simeq  \left( \frac{M_{p,hill}}{3M_*} \right)^{1/3}r_p,
 \label{eq:hill}
\end{equation}
where $M_* = 0.55 M_{\odot}$ is the stellar mass \citep{kwon11}, $r_p$ is the distance from the continuum peak to the bodies. Since the radial profile of the continuum image at $\lambda = 0.87$ mm (band 7) shows the most detailed gap structure due to the highest spatial resolutions, we use the data at band 7 (see Figure \ref{fig:fig3}a). Table \ref{table2} summarizes the resultant value of $M_{p,hill}$ at each gap. We find that the object at the innermost orbital radius is more than 80 $M_J$, which is so massive that the body can be viewed as a substellar object, rather than a planetary companion. All of the other bodies have masses approximately comparable to jovian planets. Note that while we have assumed so far that 3rd and 4th gaps are individual ones, we have also examined a possibility that these two are actually classified as one large gap. This may be implied by the radial surface brightness profile (see Figure \ref{fig:fig4}); it may show that 3rd and 4th gaps are just a sub-substructure of one large gap. If this is the case, $\Delta r_p = 16.7^{+0.4}_{-0.5}$ AU and $r_p = 71.8^{+1.4}_{-1.0}$ AU, which ends up with M$_{p,hill}$ that becomes $22\pm{3}$ $M_J$. As expected, the value is so large that the body is more likely to be a substellar object.

Based on the above discussion, the Hill radius analysis results in the value of masses that are quite large, especially for the innermost gap. This may suggest that the method itself is over-simplified, and hence leads to a significant overestimate of planetary masses. Accordingly, we now adopt a more detailed method and re-estimate the resultant masses.

There are a number of dedicated studies in which the gap structure is investigated by taking account of the balance of angular momentum transfer in gas disks under the action of tidal torque arising from planets \citep[e.g.,][]{lin79,goodman01}. Here, we follow a method derived by \citet{kanagawa15}. In the method, the gap depth (rather than the gap width) is required to estimate the planet mass, which can be written as (see their equation (7))
\begin{eqnarray}
 \label{eq:sigma}
 \frac{M_p}{M_\star} & = & 5 \times 10^{-4} \left(\frac{1}{\Sigma_p/\Sigma_0}-1 \right)^{1/2}  \\
 & \times & \left(\frac{h_p}{0.1}\right)^{5/2} \left(\frac{\alpha}{10^{-3}}\right)^{-1/2}, \nonumber
\end{eqnarray}
where $M_p$ is a planet mass, $\Sigma_p$ is the surface density at the planet orbital radius, 
$\Sigma_0$ is the surface density where no planet exists, $h_p$ is a disk aspect ratio, and $\alpha$ is viscosity, respectively.

Adopting the disk model given by \citet{kwon11} coupled with some disk parameters used by \citet{kanagawa15}, we compute the mass of hypothetical bodies that can curve the observed gaps. Table \ref{table2} tabulates the outcome. The results show that the computed value of the mass at all of the gaps becomes comparative or less than a Jupiter mass, which is consistent with the estimate done in other studies \citep{tamayo15,dipierro15}. An an additional confirmation, we have also computed the mass of bodies, following a method proposed by \citet{crida06} (The results are not shown in this paper). For this case, the gap width is needed to estimate the mass as done in the Hill radius analysis. We have found that, within the uncertainty of disk viscosity and stellar mass, the resultant values are all consistent with those summarized in Table \ref{table2}. This indicates that, while some disk quantities may be uncertain, the second approach would provides a more accurate value in the mass estimate. We therefore will use the value of masses derived from a method introduced in \citet{kanagawa15} in the following section.

\begin{table}
\begin{center}
\caption{Summary of {the gap properties and the mass of hypothetical bodies}}
\label{table2}
\begin{tabular}{c|c|ccc}
\hline
Gap & $r_p$ & $\Delta r_p$ & $M_{p,hill}$$^\dagger$ & ${M_p}^\ddagger$ \\ 
& [AU] & [AU] & [$M_J$] & [$M_J$] \\ \hline \hline 
1 & 13.5 $^{+0.4}_{-0.4}$ & 5.0 $^{+0.1}_{-0.1}$ & 88.8 $^{+5}_{-5}$ & 0.85 \\
2 & 32.4 $^{+0.6}_{-0.4}$ & 4.1 $^{+0.2}_{-0.2}$ & 3.6 $^{+0.7}_{-0.6}$  & 0.61 \\
3 & 65.2 $^{+1.3}_{-0.9}$ & 6.2 $^{+0.6}_{-0.8}$ & 1.5 $^{+0.5}_{-0.5}$ & 0.62 \\
4 & 77.2 $^{+0.8}_{-0.7}$ & 4.5 $^{+0.4}_{-0.4}$ & 0.3 $^{+0.1}_{-0.1}$ & 0.51 \\
\hline
\end{tabular}
\end{center}
$^\dagger$ the planet mass derived from Hill radius (see equation (\ref{eq:hill})) \\
$^\ddagger$ the planet mass derived from the method of \citet{kanagawa15} \\
\end{table}

\begin{table*}
\begin{minipage}{17cm}
\begin{center}
\caption{Summary of results}
\label{table3}
\begin{tabular}{c|cc|ccc|ccc}
\hline
Gap & $\Sigma_g(r_p)$ & $T(r_p)$ & $\Sigma_{g,min}(r_p)^\ast$ & $T_{min}(r_p)^\ast$ & GI & $M_{gap}$ & \multicolumn{2}{c} {Migration} \\  
& [g $\rm{cm^{-2}}$] & [K] & [g $\rm{cm^{-2}}$] & [K] & & [$M_J$] & Hill$^\dagger$  & Crida$^\ddagger$ \\ \hline \hline
1 & 45 & 30 & $1.4 \times 10^3$ & 64  & No & 0.1 & Type II & 0.14 \\
2 & 49 & 21 & $1.4 \times 10^2$ & 22 & No & 0.2 & Type II & 0.44 \\
3 & 34 & 15 & 29 & 8 & Yes & 0.36 & Type II & 0.68 \\
4 & 26 & 14 & 18 & 6 & Yes & 0.41 & Type II & 1.10 \\
\hline
\end{tabular}
\end{center}
$^\ast$ the value derived from the disk model of \citet{kwon11} \\
$^\dagger$ the results from Hill radius analysis (see equation (\ref{eq:mgap})) \\
$^\ddagger$ the results from a method proposed by \citet{crida06} \\
\end{minipage}
\end{table*}

\subsection{Formation Mechanisms}
\label{sec:formation}

We now discuss what kind of formation mechanism(s) of the bodies can be plausible to account for the observed gap structure. Currently, there are two planet formation mechanisms that are actively investigated: the core accretion (CA) and the gravitational instability (GI) scenarios. In the CA, which is the widely accepted theory of planet formation, the formation of planets divides into two phases \citep[e.g.,][]{pollack96,ida04,hasegawa14}. In the first phase, planetary cores form via the so-called runaway and oligarchic growth \citep[e.g.,][]{wetherill89,kokubo98}, while in the second phase, the surrounding gas in disks accretes on the formed cores \citep{mizuno80,stevenson82,ikoma00}. In the GI, contrastingly, planets form in gravitationally unstable disks as stars form in gravitationally unstable molecular clouds \citep[e.g.,][]{boss97,gammie01,rafikov05,cai06,kratter10,meru10}. In this case, massive disks are needed to trigger GIs there, which ends up with fragmentation of the disks and the resultant formation of massive clumps. In the following discussion, we attempt to examine the possibility that the GI might actually work in this system. This is naturally motivated by the observational results that HL Tau is currently viewed as the intermediate phase between Classes I and II, and thereby its disk may be massive enough to trigger the GI.

\begin{figure}[h]
\begin{center}
\includegraphics[scale=0.4]{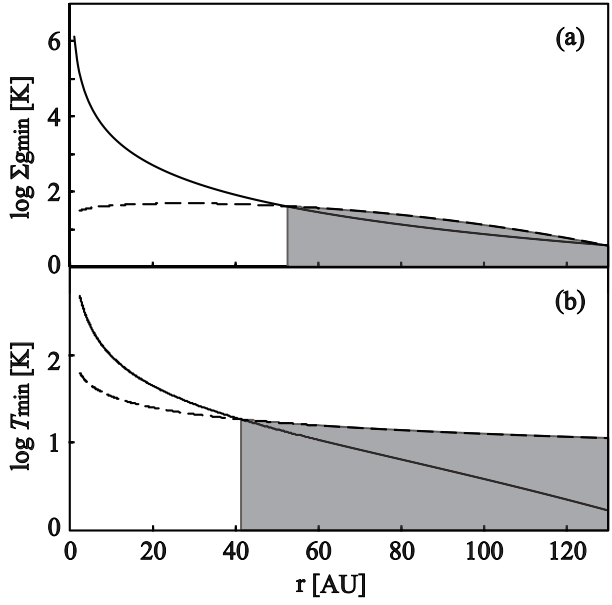}
\caption{
Panel (a) represents the surface density distribution of \citet{kwon11} model (see the solid line, also see equation (\ref{eq:sigma}))  and the minimum surface density of gas ($\Sigma_{g, min}$) that is required to trigger GIs (see the dashed curve). Panels (b) shows the radial temperature distribution given by power low function in \citet{kwon11} model (see the solid line) and the minimum disk temperature ($T_{min}$) that is required to satisfy the cooling condition of the GI (see the dashed line). The shaded area in both of the panel represents the region where the GI can be invoked.}
\label{fig:fig5}
\end{center}
\end{figure}

We first compute the minimum value of the surface density of gas ($\Sigma_{g,min}$) that is required to trigger GIs. Based on the so-called Toomre Q condition \citep{toomre64}, $\Sigma_{g,min}$ can be written as \citep{rafikov05}\footnote{Note that we assume that $\tau = \kappa \Sigma$ and include the $\Sigma-$dependence in $\tau$ to derive $\Sigma_{g,min}$.}, in the optically thick limit,
\begin{eqnarray}
\label{eq:sg_min}
\Sigma_{g,min} 
                        & \simeq  & 2.6 \times 10^{6} \mbox{ g cm}^{-2} \\
                       & \times &  \left( \frac{M_*}{M_{\odot}} \right)^{7/8} \left( \frac{r_p}{ 1 \mbox{AU}} \right)^{-21/8} 
                         \left( \frac{\kappa}{\kappa_0 } \right)^{1/4}, \nonumber
\end{eqnarray}
where $\kappa= \kappa_0\left(\nu/\nu_0\right)^\beta$ is the opacity. We set $\kappa_0 =0.01$ cm$^2$ g$^{-1}$, $\nu_0=230$ GHz and $\beta=0.729$, following the disk model of \citet{kwon11}. The resultant value of $\Sigma_{g,min}$ is summarized in Table \ref{table3}. One immediately finds that $\Sigma_g$ at $r =$ 65.2 AU and $r=$ 77.2 AU exceeds the value of $\Sigma_{g,min}$, indicating that the minimum condition of the GI is satisfied at these two positions. To further visualized this trend, Figure \ref{fig:fig5} a shows the region in which GIs can operate (see the hatched region).

However, we cannot fully conclude yet that the GI can really operate there. This is because it is important to examine the cooling condition that assures that disks fragment into self-gravitating clumps \citep[e.g.,][]{gammie01}. To proceed, we calculate the minimum disk temperature ($T_{min}(r)$) that is required to satisfy the cooling condition. The actual formula can be given as \citep{rafikov05}, in the optically thick limit, 
\begin{eqnarray}
T_{min}(r) 
& \simeq  & 1.4 \times 10^{3} \mbox{ K} \left( \frac{r}{ 1 \mbox{ AU}} \right)^{-6/5}  \\
& \times &  \left( \frac{M_*}{ 1 M_{\odot}} \right)^{2/5} 
                  \left( \frac{\kappa}{\kappa_0 } \right)^{2/5} \left( \frac{\Sigma_g}{ 10 \mbox{ g cm}^{-2} } \right)^{2/5}. \nonumber
\end{eqnarray}
The resultant value of $T_{min}(r_p)$ is summarized in Table \ref{table3}. We find that $\Sigma_{g}(r_p)$ and $T(r_p)$ at $r=65.2$ AU and $r=77.2$ AU exceeds $\Sigma_{g,min}$ and $T_{min}(r_p)$. As done in Figure \ref{fig:fig5} a, the region where GIs can be feasible in terms of $T_{min}(r_p)$ is denoted by the hatched region in Figure \ref{fig:fig5} b. Our analysis therefore suggests that, when the observed gaps originate from the presence of unseen bodies, the GI is very likely to be a mechanism for the gaps at $r=65.2$ and $77.2$ AU. Note that our finding is consistent with the results of \cite{kwon11}, which indicate that the HL Tau disk may be gravitationally unstable in the outer region, especially the region between 50 AU and 100 AU in radius. 

In the case of HL Tau, therefore, GIs can occur in the region beyond $r \sim$ 52 AU, where both of the conditions are simultaneously met.

\subsection{Origins of hypothetical bodies at the gaps}
\label{migration}

As discussed above, the gaps located in the outer part of the disk may be related to GIs, which can form massive clumps, possibly the progenitors of planets. How about the gaps located in the inner part of the disk (at $r=13.5$ and $32.4$ AU)? Is there any possibility that an origin of these gaps can be discussed in the same framework as above? To this end, we focus on planetary migration that is induced mainly by the disk gas. Here, we attempt to discuss how all of the observed gaps can be established, coupling GIs with planetary migration.

Planetary migration arises from tidal resonant interactions between massive bodies and the surrounding gas disks, which leads to the efficient angular momentum transfer between them \citep[e.g.,][for a review]{kley12}. There are mainly two modes in migration that are classified by the mass of migrating bodies: type I and type II \citep[e.g.,][]{ward97,nelson00,tanaka02,paardekooper09,hasegawa13}. Type I migration is effective for low-mass bodies such as cores of gas giants and terrestrial planets, and the direction of the migration is very sensitive to the disk structure such as the surface density and the temperature. Type II migration becomes important for massive bodies such as Jovian planets in our solar system, and its direction is generally regulated by the bulk gas flow in the disks, that is, spiraling toward the central star.

We first determine which type of migration can take place in the HL Tau system. This can be done by estimating the so-called gap opening mass ($M_{gap}$) above which planets are sufficiently massive to open up a gap in their gas disk via disk-planet interactions, and hence the planets will undergo type II migration. In principle, $M_{gap}$ can be formulated either by the Hill radius analysis or by a disk viscosity analysis \citep[e.g.,][]{crida06,hasegawa14}. We first adopt the Hill radius analysis to compute $M_{gap}$, which can be written as
\begin{eqnarray}
\label{eq:mgap}
M_{gap} & \simeq & 2.7 M_{\oplus}  \\ 
         & \times & \left( \frac{r}{ 1 \mbox{ AU}} \right)^{3/2}\left( \frac{M_*}{1 M_{\odot}} \right)^{-1/2}\left( \frac{T}{100 \mbox{K}} \right)^{3/2}. \nonumber
\end{eqnarray}
Note that the HL Tau disk would have a high value of disk accretion rates, equivalently a high value of the disk turbulence. As a result, our estimate may become conservative in the sense that the Hill radius analysis would provide a lower value of $M_{gap}$. The resultant value of $M_{gap}$ is summarized in Table \ref{table3}. We find that type II migration would be a dominant mode of migration at all of the gaps. 

We now can examine how valid the Hill radius analysis is by adopting a disk viscosity analysis. We relay on a criterion of Type II migration derived from \citet{crida06}, which can be written as (see their equation (15))
\begin{eqnarray}
\label{eq:mgap}
R=\frac{3}{4}\frac{H}{R_H}+\frac{50}{\hat{q}Re} \lesssim 1,   
\end{eqnarray}
where $H$ is the pressure scale height of the disk, $R_H$ is the Hill radius, $\hat{q}$ is the mass ratio of the planet to the central star $M_p$/$M_\star$, and $Re$ is the Reynolds number $\hat{r}^2_p$$\Omega_p$/$\nu$, where $\hat{r}_p$ is the distance to the planet from the central star, $\Omega_p$ is the angular velocity of the planet, and $\nu$ is the disk viscosity, respectively.

Based on the criterion, planets can undergo Type II migration when the value is less than unity, and they can experience Type I migration when it becomes larger than unity. Table \ref{table3} summarizes the resultant value. We find that all of the hypothetical bodies may undergo Type II migration, except the outermost body. The resultant $R$ for the outermost one marginally exceeds unity. In addition, it can naturally be expected that the disk viscous parameter ($\alpha$) would have a relatively large uncertainty, which can end up with an overestimate of $R$. Thus, it can be concluded that both the approaches lead to a roughly consistent consequence, which suggests that type II migration can be effective for hypothetical bodies. 

If Type II migration is the case, we can develop an intriguing conjecture as follows: the hypothetical bodies that might play some role in opening up the two inner gaps ($r=13.5$ AU and $r=32.4$ AU) might have formed originally in the outer part of the disk where GIs can operate. Then, the bodies might subsequently have undergone type II migration. Since the direction of type II migration is inward for most of the cases, they would eventually have arrived at the current locations. In other words, the outer part of the disk would act as the main site of forming all of the massive bodies, and all of the observed gaps in the HL Tau system might be established by the following inward migration of the bodies.

The unprecedented high spatial resolution images of HL Tau derived from ALMA observations have a substantial potential of revolutionizing our picture of planet formation. As discussed above, the observed multiple gap structure may be a consequence of the coupling of GIs with inward planetary migration. Nonetheless, it is obvious that further observational and theoretical studies are needed to examine our conjecture.  For instance, the current image quality can be further improved by conducting long-baseline observations. Moreover, optically thin line observations such as $^{13}$C$^{18}$O or deuterium molecules are demanded to investigate the gas distribution near the mid-plane of the disk. We also cannot fully exclude the hypothesis that the estimated Jovian bodies in the analysis might form via CA. 

\section{Conclusion}
\label{conclusion}

We re-processed  ALMA science verification archive data of HL Tau and as a result, a better continuum image in particular at $\lambda = 0.87$ mm was obtained. The radial surface brightness profiles derived from the new images precisely display a clear outline of gap structures. Assuming the observed gaps are opened up by unseen bodies, we estimate the mass of such bodies by following 2 approaches; a simple Hill radius analysis and a more elaborated approach developed from the angular momentum transfer analysis. In the Hill radius analysis, the estimated planet mass results in a large value in particular at the innermost gap, indicating that the analysis is too simplified. In the angular momentum transfer analysis, we apply \citet{kanagawa15} model and as a result, the planet masses are comparable or less than Jovian planet mass. We also confirmed that the same results can be obtained by \citet{crida06} model. 

By evaluating Toomre's GI condition and cooling effect, we found that the GI can operate in the region beyond 50 AU from the central star. We also show the bodies are massive enough to open up the gas disk and hence can undergo inward type II migration. Combining GIs with inward migration, we conjecture that all of the observed gaps may be a consequence of bodies that might have originally formed at the outer part of the disk, and have subsequently migrated to the current locations. The optically thin gas observations with high spatial resolution that can trace the gas around the mid-plane would bring us  to fully understand the HL Tau images. In our subsequent paper, we will undertake more detailed radiative transfer modeling and perform a careful comparison with the observations, which will enable us to develop a more quantitative analysis of our conjecture.

\acknowledgments
This paper makes use of the following ALMA data: ADS/JAO.ALMA$\sharp$2011.0.00015.SV. ALMA is a partnership of ESO (representing its member states), NSF (USA) and NINS (Japan), together with NRC (Canada), NSC and ASIAA (Taiwan), and KASI (Republic of Korea), in cooperation with the Republic of Chile. The Joint ALMA Observatory is operated by ESO, AUI/NRAO and NAOJ. Y.H. is supported by EACOA Fellowship that is supported by East Asia Core Observatories Association which consists of the Academia Sinica Institute of Astronomy and Astrophysics, the National Astronomical Observatory of Japan, the National Astronomical Observatory of China, and the Korea Astronomy and Space Science Institute. We would like to thank Crystal L. Brogan and Takashi Tsukagoshi for useful comments on imaging and Kazuhito Kanagawa for productive discussion on planet mass estimation. We are also grateful to the anonymous referee who helped improve the manuscript.

\clearpage
\clearpage

\clearpage

\clearpage

\clearpage

\clearpage

\clearpage

\begin{thebibliography}{}
\expandafter\ifx\csname natexlab\endcsname\relax\def\natexlab#1{#1}\fi
\bibitem[Akiyama et al.(2015)]{akiyama15} Akiyama, E., Muto, T., Kusakabe, N., et al.\ 2015, \apjl, 802, L17 
\bibitem[ALMA Partnership et al.(2015)]{partnership} ALMA Partnership, Brogan, C.~L., P{\'e}rez, L.~M., et al.\ 2015, \apjl, 808, L3 
\bibitem[Andrews et al.(2011)]{andrews11} Andrews, S.~M., Wilner, D.~J., Espaillat, C., et al.\ 2011, \apj, 732, 42
\bibitem[Boss(1997)]{boss97} Boss, A.~P.\ 1997, Science, 276, 1836
\bibitem[Cai et al.(2006)]{cai06} Cai, K., Durisen, R.~H., Michael, S., et al.\ 2006, \apjl, 636, L149 
\bibitem[Crida et al.(2006)]{crida06} Crida, A., Morbidelli, A., \& Masset, F.\ 2006, \icarus, 181, 587 
\bibitem[Debes et al.(2013)]{debes13} Debes, J.~H., Jang-Condell, H., Weinberger, A.~J., Roberge, A., \& Schneider, G.\ 2013, \apj, 771, 45 
\bibitem[Dipierro et al.(2015)]{dipierro15} Dipierro, G., Price, D., Laibe, G., et al.\ 2015, \mnras, 453, L73 
\bibitem[Dullemond et al.(2001)]{dullemond01} Dullemond, C.~P., Dominik, C., \& Natta, A.\ 2001, \apj, 560, 957 
\bibitem[Elias(1978)]{elias78} Elias, J.~H.\ 1978, \apj, 224, 857 
\bibitem[Fukagawa et al.(2013)]{fukagawa13} Fukagawa, M., Tsukagoshi, T., Momose, M., et al.\ 2013, \pasj, 65, L14 
\bibitem[Gammie(2001)]{gammie01} Gammie, C.~F.\ 2001, \apj, 553, 174 
\bibitem[Grady et al.(2013)]{grady13} Grady, C.~A., Muto, T., Hashimoto, J., et al.\ 2013, \apj, 762, 48 
\bibitem[Goodman \& Rafikov(2001)]{goodman01} Goodman, J., \& Rafikov, R.~R.\ 2001, \apj, 552, 793 
\bibitem[{{Hasegawa} \& {Ida}(2013)}]{hasegawa13}{Hasegawa}, Y. \& {Ida}, S. 2013, \apj, 774, 146
\bibitem[Hasegawa \& Pudritz(2014)]{hasegawa14} Hasegawa, Y., \& Pudritz, R.~E.\ 2014, \apj, 794, 25 
\bibitem[Hashimoto et al.(2012)]{hashimoto12} Hashimoto, J., Dong, R., Kudo, T., et al.\ 2012, \apjl, 758, L19 
\bibitem[Ida \& Lin(2004)]{ida04} Ida, S., \& Lin, D.~N.~C.\ 2004, \apj, 604, 388 
\bibitem[Ikoma et al.(2000)]{ikoma00} Ikoma, M., Nakazawa, K., \& Emori, H.\ 2000, \apj, 537, 1013 
\bibitem[Isella et al.(2013)]{isella13} Isella, A., P{\'e}rez, L.~M., Carpenter, J.~M., et al.\ 2013, \apj, 775, 30 
\bibitem[Kanagawa et al.(2015)]{kanagawa15} Kanagawa, K.~D., Muto, T., Tanaka, H., et al.\ 2015, \apjl, 806, L15  
\bibitem[Kenyon et al.(1994)]{kenyon94} Kenyon, S.~J., Dobrzycka, D., \& Hartmann, L.\ 1994, \aj, 108, 1872 
\bibitem[{Kley \& Nelson(2012)}]{kley12} Kley, W. \& Nelson, R.~P. 2012, \araa, 50, 211
\bibitem[Kokubo \& Ida(1998)]{kokubo98} Kokubo, E., \& Ida, S.\ 1998, \icarus, 131, 171 
\bibitem[Kratter et al.(2010)]{kratter10} Kratter, K.~M., Murray-Clay, R.~A., \& Youdin, A.~N.\ 2010, \apj, 710, 1375 
\bibitem[Kwon et al.(2011)]{kwon11} Kwon, W., Looney, L.~W., \& Mundy, L.~G.\ 2011, \apj, 741, 3 
\bibitem[Lin \& Papaloizou(1979)]{lin79} Lin, D.~N.~C., \& Papaloizou, J.\ 1979, \mnras, 186, 799 
\bibitem[Meru \& Bate(2010)]{meru10} Meru, F., \& Bate, M.~R.\ 2010, \mnras, 406, 2279 
\bibitem[Men'shchikov et al.(1999)]{menshchikov99} Men'shchikov, A.~B., Henning, T., \& Fischer, O.\ 1999, \apj, 519, 257
\bibitem[Mizuno(1980)]{mizuno80} Mizuno, H.\ 1980, Progress of Theoretical Physics, 64, 544 
\bibitem[Monin et al.(1996)]{monin96} Monin, J.-L., Pudritz, R.~E., \& Lazareff, B.\ 1996, \aap, 305, 572 
\bibitem[{Nelson {et~al.}(2000)Nelson, Papaloizou, Masset, \& Kley}]{nelson00} Nelson, R.~P., Papaloizou, J. C.~B., Masset, F., \& Kley, W. 2000, \mnras, 318, 18
\bibitem[Hayashi et al.(1993)]{hayashi93} Hayashi, M., Ohashi, N., \& Miyama, S.~M.\ 1993, \apjl, 418, L71
\bibitem[{Paardekooper {et~al.}(2010)Paardekooper, Baruteau, Crida, \& Kley}]{paardekooper09} Paardekooper, S.-J., Baruteau, C., Crida, A., \& Kley, W. 2010, \mnras, 401, 1950 
\bibitem[Pollack et al.(1996)]{pollack96} Pollack, J.~B., Hubickyj, O., Bodenheimer, P., et al.\ 1996, \icarus, 124, 62 
\bibitem[Pringle(1981)]{pringle81} Pringle, J.~E.\ 1981, \araa, 19, 137 
\bibitem[Rafikov(2005)]{rafikov05} Rafikov, R.~R.\ 2005, \apjl, 621, L69 
\bibitem[Sargent \& Beckwith(1991)]{sargent91} Sargent, A.~I., \& Beckwith, S.~V.~W.\ 1991, \apjl, 382, L31
\bibitem[Stevenson(1982)]{stevenson82} Stevenson, D.~J.\ 1982, \planss, 30, 755 
\bibitem[Takahashi \& Inutsuka(2014)]{takahashi14} Takahashi, S.~Z., \& Inutsuka, S.-i.\ 2014, \apj, 794, 55 
\bibitem[Tamayo et al.(2015)]{tamayo15} Tamayo, D., Triaud, A.~H.~M.~J., Menou, K., \& Rein, H.\ 2015, \apj, 805, 100 
\bibitem[{Tanaka {et~al.}(2002)Tanaka, Takeuchi, \& Ward}]{tanaka02} Tanaka, H., Takeuchi, T., \& Ward, W.~R. 2002, \apj, 565, 1257
\bibitem[Thalmann et al.(2010)]{thalmann10} Thalmann, C., Grady, C.~A., Goto, M., et al.\ 2010, \apjl, 718, L87 
\bibitem[Thalmann et al.(2014)]{thalmann14} Thalmann, C., Mulders, G.~D., Hodapp, K., et al.\ 2014, \aap, 566, A51 
\bibitem[Toomre(1964)]{toomre64} Toomre, A.\ 1964, \apj, 139, 1217
\bibitem[{Ward(1997)}]{ward97} Ward, W.~R. 1997, \icarus, 126, 261
\bibitem[Ward(2000)]{ward00} Ward, W.~R.\ 2000, Origin of the Earth and Moon, 75 
\bibitem[Wetherill \& Stewart(1989)]{wetherill89} Wetherill, G.~W., \& Stewart, G.~R.\ 1989, \icarus, 77, 330 
\bibitem[Youdin(2011)]{youdin11} Youdin, A.~N.\ 2011, \apj, 731, 99
\bibitem[Zhang et al.(2015)]{zhang15} Zhang, K., Blake, G.~A., \& Bergin, E.~A.\ 2015, arXiv:1505.00882 

\end{thebibliography}
\end{document}